\documentstyle [12pt,epsfig]{article}
\begin{document}
\begin{center}
{\Large \bf New Results on Multiplicity, Identified Particles and
Spin Measurements in {\boldmath $e^+e^-$} collisions} \\ ~ \\
{\large {\it Vasilii Nomokonov} (Helsinki Institute of Physics)} \\ 
{\large on behalf of the DELPHI collaboration} \\ ~ \\
\end{center}
\begin{center}
Abstract 
\end{center}
The analyses of recent LEP data provide
us with a rich ground for pQCD and LPHD tests. A number of non-trivial
effects predicted in the framework of MLLA are experimentally verified.
Recent measurements also enrich our
knowledge about the hadronization process.\\

\section{Introduction}
Recent LEP results on multiparticle production in hadronic jets enable
further important tests of QCD. One of the most 
interesting and intriguing outcomes of the pQCD calculations
is the prediction of so-called colour coherence effects
\cite{deadcone}.

A number of analyses presented here demonstrate that there is an
agreement of the data with the QCD MLLA
predictions and provide further support for the concept of
local parton hadron duality (LPHD) \cite{LPHD}.

There are also new results in the physics of large distances, where 
the processes
not calculable within the pQCD provide us with abundant information
needed for tuning hadronization model parameters.

\section{Testing analytical predictions of Perturbative QCD}


Inclusive charged hadron distributions measured by DELPHI 
at 189 GeV \cite{225} were presented as a function of the
variables: rapidity, $\xi_p=ln(1/x_p)$, momentum and transverse momentum.
The $\xi_p$ distribution demonstrates the so-called ``hump backed'' 
behaviour predicted for partons in the framework of the Modified 
Leading Logarithmic Approximation (MLLA) \cite{LPHD}. 
The simultaneous fit to the
$\xi_p$ distribution with a Fong-Webber distorted Gaussian \cite{FongWebber}
at different 
energies including the present measurement at 189 GeV show very good
agreement of the data and the prediction ($\chi^2/dof=99.6/97$)
giving support for the LPHD hypothesis.

MLLA also provides a definite prediction for the energy evolution of 
the maximum of the $\xi$ distribution, $\xi^*$.
As hadronization and resonance decays are expected to act similarly at
different centre-of-mass energies,
the energy evolution of $\xi^*$ is expected to be
less sensitive to nonperturbative effects.
The $\xi^*$ values entering in the analysis were determined by fitting a 
distorted Gaussian with the parameters given by the Fong-Webber 
parametrisation.
For the 189 GeV  data one obtains $\xi^* = 4.157 \pm 0.030$.
A fit of the MLLA prediction again demonstrates good agreement while
ruling out a phase space expectation and the DLA prediction.

The energy evolution of the momentum distribution is well described 
by the fragmentation model. An interesting observed feature is the 
approximate $E_{CM}$ independence of hadron production at very small momenta 
$p<1$~GeV. This has been explained in \cite{Phys_LettB394},\cite{LPHD} 
to be due to
the coherent emission of long wavelength gluons by the total colour current
which is independent of the internal jet structure and is 
conserved under parton splittings.
Therefore, low-energy gluon emission is expected to be almost independent on
the number of hard gluons radiated and hence of the centre-of-mass energy.
Provided the LPHD hypothesis is correct,
the number of produced hadrons at small momenta is approximately constant.


A sample of 2.2 million hadronic $Z$ decays, selected from the data
recorded by the Delphi detector at LEP during 1994-1995
was used for a precise measurement of inclusive distributions of
$\pi^+$, $K^+$ and $p$ and their anti-particles in gluon and
quark jets \cite{146}. As observed for inclusive charged particles,
the production spectra of the individual identified particles were found to be
softer in gluon jets compared to the quark jets, with a higher multiplicity
in gluon jets. A significant enhancement of protons 
in gluon jets is observed. The ratio of the average multiplicity
in $g$ jets with respect to $q$ jets
was found for all identified particles to be consistent with the
ratio measured for all charged particles. The normalised ratio for protons
in Y events was measured to be:
$$
 R_{p} = 1.205 \pm 0.041 ,
$$
which differs significantly from unity.

The maxima, $\xi^*$, of the $\xi$-distributions for kaons in gluon and quark
jets are observed to be different.


A particularly nice illustration of the phenomenon of QCD coherence and 
a test of LPHD 
was obtained by DELPHI using symmetric 3-Jet Events \cite{145}. It is
known that soft radiation is sensitive to the total colour flow in the 
underlying hard partonic structure. Let's consider, for example, two extreme
two-jet topologies of a $q\bar{q}g$ event. If the gluon is
collinear to one of the quarks, the colour flow in the event 
will be identical to the $q\bar{q}$ case, whereas 
if the gluon exactly recoils with respect to the two quarks, the
colour flow will correspond to that of a $gg$ event. In the latter case
the soft radiation at large angle is expected to be increased by the 
colour   factor ratio $C_A/C_F$ as compared to the $q\bar{q}$ case. 
The evolution between those extreme cases has been calculated as a function 
of the opening angles between the jets \cite{Phys_LettB394}.
Thus, the charged hadron multiplicity in a cone perpendicular
to the event plane of symmetric three-jet events was determined as 
a function of an inter-jet angle for the data collected at the Z
resonance. A clear dependence of 
the multiplicity on the opening angle was observed and appears to 
be in agreement with QCD predictions \cite{YA2},\cite{Phys_LettB394}.


An interesting example of the intra-jet QCD coherence is the restriction of
forward gluon emission for heavy quarks. A calculation in the
framework of MLLA predicts the following angular
distribution of gluon emission \cite{deadcone}:
$$ \frac{dn}{d\theta^2}\sim \theta^2/(\theta^2+\theta^2_{min})^2,~~
\theta_{min}\equiv \frac{m_Q}{E}$$
where $m_Q$ and $E$ are a quark mass and energy, respectively.
Provided LPHD holds, the effect should be seen, for example, 
in a comparison of the primary-particle angular distribution for $b\bar{b}$
and $q\bar{q}$ (where q denotes u,d or s quark) Z decays. Delphi
has presented preliminary results that show a difference in this
behaviour. Hadrons containing the original quark or originating from 
the decay of such particle are carefully excluded from 
consideration.


The phenomenon of colour coherence is not only a subject for tests,
but also may be used as a tool for reconstruction of the event colour 
structure on an event-by-event basis.
The idea presented in \cite{method} is based on the fact that 
soft particles do not originate from a particular parton but
rather their production depends on the whole colour topology of the event.
Therefore, it is possible to define a way to
estimate the colour connection strength between the partons by analysing
the behaviour of the soft particles. The method 
can then be used for parton identification.
The proposed algorithm is described below: first, fast particles are used
to reconstruct cluster directions and then to define
a weight $w_{ij}$ that a particle $i$ may be connected to a cluster $j$ with:
$$ w_{ij} = \frac{C_i}{k_{ij}^2}$$ 
where $k_{ij}^2 = 2 E_i^2 (1-cos\Theta_{ij})$, with normalisation 
$\Sigma_{j} w_{ij} = 1$. Then each particle with $w_{ij} < 0.95$ is 
assigned to the cluster pair $kl$ for
which the sum $w_{ikl} = w_{ik} + w_{il}$ is maximal.
Then parton pair connectedness is defined as
$$ W_{kl} = C_{kl} \Sigma_i g(E_i)(w_{ik} + w_{il})$$
where $i$ runs through all the particles assigned to the clusters $kl$. 

The method was tested by the Delphi collaboration by using
double b-tagged 3-jet Mercedes events collected at the Z pole.
The gluon jet is known in these kind of events as one which is not
b-tagged. On the other hand, the gluon should have
two colour connections and thus the colour connectedness $W_{kl}$
of the $b\bar{b}$ pair must have the smallest value in any given event. 
Matching
the two gives the purity of the method which is found to be above $60\%$ and
could be improved by requiring the smallest colour connection coefficient
to be below a predetermined threshold value.

The method with various modifications could be used 
for identifying colour connections in
numerous applications (pairing, background rejection).


One of the approaches to study a parton shower cascade 
is to employ the multiplicity moments technique.
The oscillations in the ratio of the cumulant factorial to the 
factorial charged 
particle multiplicity moments in Z Decays is known to show a 
quasi-oscillatory behaviour when plotted versus the order of the moment,
as was observed by the SLD collaboration some time ago \cite{SLD}.
This peculiarity is also predicted by the NNLLA of perturbative QCD within
the LPHD framework \cite{Dremin}. 

However, using the jet multiplicity distributions obtained from
the Cambridge jet algorithm, in order to vary the dependence on the LPHD 
hypothesis, the L3 collaboration found \cite{276} 
that the oscillations appear only for non-perturbative 
energy scales, namely $\leq 100$ MeV. From this conclusion it follows
that the observed oscillations are unrelated with the behaviour predicted by
the NNLLA perturbative QCD calculations.


Another challenging way to study the cascade is to measure
multiplicity fluctuations in rings around the jet axis and in
off-axis cones. The DELPHI collaboration performed the measurement 
\cite{222} and compared them with analytical perturbative QCD
calculations for the corresponding multiparton system, using the concept
of LPHD. Some qualitative features were confirmed 
by the data but substantial quantitative deviations are observed.


\section{Fragmentation physics}


Our knowledge about the hadronization process has been significantly
enriched by recent measurements at LEP.

Thus, results on the production of the $\Lambda(1520)$ are presented, 
as obtained from hadronic Z decays recorded by DELPHI \cite{147}.
The $\Lambda(1520)$ scaled momentum ($x_p$) spectrum is determined. 
The relative importance of $\Lambda(1520)$ production increases with $x_p$
similarly to that of orbitally excited mesons.
It is shown that the $\Lambda(1520)$ primarily originates from fragmentation 
and not from heavy particle ($b, c$) decays.
The large $\Lambda(1520)$  production rate 
$N_{\Lambda(1520)}/N_Z = 0.030 \pm 0.004 \pm  0.005$
suggest that many stable baryons descend from 
orbitally excited baryonic states.


The OPAL collaboration measured
the helicity density matrix elements $\rho_{00}$ of
$\rho(770)^{\pm}$ and $\omega(782)$ mesons
produced in Z$^0$ decays \cite{63}. Over the measured meson energy range,
the values are compatible with 1/3,
corresponding to a statistical mix of helicity $-1$, 0 and 1 states.
For the highest accessible scaled energy range 0.3 $<$ $x_E$ $<$ 0.6,
the measured $\rho_{00}$ values of the $\rho^{\pm}$
and the $\omega$ are 0.373~$\pm$~0.052 and
0.142~$\pm$~0.114, respectively.


The ALEPH collaboration performed an extensive study
of the production rates and the inclusive cross 
sections of the isovector meson $\pi^0$, the isoscalar mesons $\eta$ and 
$\eta^{\prime}(958)$, the strange meson $\mathrm{K^0_S}$ and the $\Lambda$ 
baryon. This was done as function of scaled energy (momentum) 
in hadronic events, 
two-jet events and each jet of three-jet events from hadronic $\mathrm{Z}$
decays and compared the results to Monte Carlo models \cite{394}. 
The JETSET modelling of the gluon fragmentation into isoscalar mesons is found 
to be in agreement with the experimental results for the measured region.
HERWIG fails to describe the $\mathrm{K^0_S}$ spectra in gluon-enriched jets 
and the $\Lambda$ spectra in quark jets.


An interesting idea which helped to understand the production rates of 
light-flavour hadrons was proposed by P.Chliapnikov \cite{561}: 
the difference between the production rates of hadrons composed of the
same quarks and belonging to the different SU(3) multiplets 
but the same SU(6) multiplet is essentially determined by the hyperfine mass
splitting. This trend shows up when the direct production rates are plotted
versus the sum of the constituent quark masses $\Sigma_i(m_q)_i$. In this
case the vector-to-pseudoscalar and decuplet-to-octet suppressions
are found to be the same. In the 
proposed scenario the strangeness suppression factor, 
$\lambda = 0.295 \pm 0.006$, is the same for mesons and baryons and 
related to the difference in the constituent quark masses, 
$\lambda = e^{-(m_s-\hat{m})/T}$, where $\hat{m} = m_u = m_d$, 
and the temperature $T = 142.4 \pm 1.8$~MeV/$c^2$. 



\begin{thebibliography}{99}
\bibitem{deadcone} For a recent review see 
Khoze V A, Ochs W 1997, {\it Int. J. of Mod. Phys.}
                   A {\bf 12} N17
\bibitem{LPHD} Azimov Ya et al. 1985, {\it Z.Phys.} C {\bf 27} 65; 
               Azimov Ya et al. 1986, {\it Z.Phys.} C {\bf 31} 213
\bibitem{225} DELPHI collaboration, EPS HEP99, contributed paper \#1\_225
\bibitem{FongWebber} Fong C P and Webber B R 1989, {\it Phys. Lett.}
                     B {\bf 229} 289
\bibitem{Phys_LettB394} Khoze V, Lupia S and Ochs W 1997, 
         {\it Phys. Lett.} B {\bf 394} 179
\bibitem{146} DELPHI collaboration, EPS HEP99, contributed paper \#3\_146
\bibitem{145} DELPHI collaboration, EPS HEP99, contributed paper \#1\_145
\bibitem{YA2} Azimov Ya et al. 1985 {\it Phys. Lett.} B {\bf 165} 147
\bibitem{method} Orava R et al. 1998, {\it Eur. Phys. J.} C {\bf 5}
                 471 
\bibitem{SLD} Abe K et al. 1996, {\it Phys. Lett.} B {\bf 371} 149
\bibitem{Dremin} Dremin I M and Nechitailo V A 1993, 
                 {\it JETP Lett.} {\bf 58} 881
\bibitem{276} L3 collaboration, EPS HEP99, contributed paper \#1\_276
\bibitem{222} DELPHI collaboration, EPS HEP99, contributed paper \#1\_222
\bibitem{147} DELPHI collaboration, EPS HEP99, contributed paper \#3\_147
\bibitem{63}  OPAL collaboration, EPS HEP99, contributed paper \#3\_63
\bibitem{394} ALEPHI collaboration, EPS HEP99, contributed paper \#1\_394
\bibitem{561} DELPHI collaboration, EPS HEP99, contributed paper \#3\_561
\end{thebibliography}
\end{document}